# DeepGene Transformer: Transformers for the Gene Expression based Classification of Cancer Subtypes

Anwar Khan, Boreom Lee*, *Member, IEEE*

***Abstract*— Adenocarcinoma and squamous cell carcinoma constitute approximately 40% and 30% of all lung cancer subtypes, respectively, and display a wide range of heterogeneity in terms of clinical and molecular responses to therapy. Molecular subtyping has enabled the use of precision medicine to overcome these challenges and provide significant biological insights to predict prognosis and improve clinical decision making. Over the past decade, conventional ML algorithms and DL based CNNs have been widely espoused for the classification of cancer subtypes from gene expression datasets. However, these methods have not completely utilized the advantages of high throughput next-generation sequencing technologies and seem to be potentially biased toward the identification of cancer markers. Recently proposed transformer-based architectures that leverage the self-attention mechanism encode high throughput gene expressions and learn representations that are computationally complex and parametrically expensive. However, compared to the datasets for NLP application, gene expression consists of several hundreds of thousands of genes from a limited number of observations, making it difficult to efficiently train transformers for bioinformatics applications. To this end, we propose an end-to-end deep learning approach, DeepGene Transformer, which addresses the complexity of high-dimensional gene expression with a multi-head self-attention module by identifying relevant biomarkers across multiple cancer subtypes without requiring feature selection as a prerequisite for the current classification algorithms. The proposed architecture achieved an overall improved performance for all evaluation metrics, and had fewer misclassification errors than the commonly used traditional classification algorithms. The classification results show that DeepGene Transformer can be an efficient approach for classifying cancer subtypes, indicating that any improvement in deep learning models in computational biologist can be reflected well in this domain as well.**

## 1. Introduction

Genes, a unit of inheritance, are responsible for storing genetic information in all living organisms, and their expression in a particular cell determines the functionality of that cell. Therefore, any type of gene mutation, single or multiple, can lead to a dysregulation in the gene expression, and this is broadly termed as genetic disorder. Cancer is a type of a genetic disorder wherein proto-oncogenes are mutated into an oncogene (cancer-causing gene) state. According to the World Health Organization [1], approximately 18.1 million cancer cases were reported in 2018. Of these, 9.6 million patients died because of the top 10 cancer types. Based on the statistics published in 2020, WHO has predicted that the rate of intercontinental cancer incidence will reach approximately 37 million by 2040. In 2018, lung cancer was the most frequently diagnosed cancer, accounting for 11.6% of all cases. Lung carcinoma, another leading cause of cancer-related death, accounted for 18.4% of all deaths. To delineate lung cancer malignancy, it is necessary to differentiate between lung cancer subtypes, such as adenocarcinoma and squamous cell carcinoma. These two communal subtypes are critical for selecting the quintessential treatment against specific genetic alterations [2, 3] in patients with carcinoma [4].

To initiate cancer therapy, it is necessary to ensure early diagnosis using precise tools. Substantial cancer research has been conducted to minimize the impact of therapy on non-cancerous cells and identify biomarkers for early prognosis. The advent of genome-wide next-generation sequencing techniques has revolutionized research in bioinformatics, precision medicine, and drug discovery [5]. Next-generation sequencing technology presents abundant data in the form of high-dimensional imbalanced gene expression levels from a low number of patient samples, from which a few genes that play a significant role in cancer can be extracted. Because cancer is a multifactorial disorder driven by genomic aberrations [6], classification of tumor-specific gene expression samples allows the study of clinical features and prognostic factors associated with tumor-specific survival or disease progression over time [7].

Accurate prediction of cancer subtypes, disease-specific survival, or disease progression estimation remains a challenge for computational biologists. Machine learning (ML) approaches are now considered inappropriate for analyzing high throughput datasets, and have made carcinoma classification a non-trivial task. A major shortcoming of conventional ML methodologies is that they require a pre-configured arrangement of unprocessed gene expression data into an organized structured form [8]. The inability of traditional ML algorithms to utilize unstructured input information has curbed their effectiveness for cancer classification [9]. The ability of deep learning (DL) algorithms to learn latent representations over unfabricated unstructured data are an advantage over conventional ML algorithms. DL has revolutionized the field of computer vision, more specifically image classification [10], visual event recognition [11], and machine translation [12], and is now considered as an effective diagnostic pipeline for disease classification in medical imaging [13] and computational biology [14].

To date, considerable investigations have been conducted on cancerous and normal samples to predict prognosis and diagnosis, and determine treatment strategies for cancer. The primary objective is to classify tumor subtypes and determine influential genes in each subtype. This can assist in developing a unified approach to recognize tumor subtypes or identify early-stage cancer.

### *1.1 Classification using conventional ML*

At this stage, different ML algorithms are trained using a subset of features extracted from the feature selection algorithms. Most proposed classification models extend from statistical tests to conventional ML algorithms such as naïve Bayes (NGB), *k*-nearest neighbor (KNN), random forest (RF), logistic regression (LR), and support vector machine (SVM). A recent study has evaluated several classification algorithms, including SVM, KNN, NGB, and decision tree for breast and lung cancer subtype classification [15, 16], and found that on utilizing features extracted at various thresholds, SVM classified cancer subtypes more precisely and had fewer prediction errors than the other algorithms.

Although few algorithms work well for binary classification problems, they are not extensible to certain trivial, imbalanced and multiclass classification tasks. In the abovementioned algorithms, only the classification accuracy in gene classification was explored. Further, several tradeoffs have been found, including maintaining accuracy versus assuring generalization, regulating complexity versus enhancing classifier performance, and improving performance versus memory requirements, which calls into question the efficiency of the cancer classification algorithms [17].

### *1.2 DL and gene expression*

Owing to the promising outcomes of CNNs in computer vision, the medical research community has shifted its focus toward DL and deploying neural networks for gene expression analysis. Several CNNs have been introduced in a study [18], in which unstructured gene expression inputs are used to categorize data as cancerous or normal. The authors built three CNN models, namely, 1D-CNN, 2D-Vanilla-CNN, and 2D-expression profiles, from a total of 10,340 cancer samples and 713 matched normal tissue samples from The Cancer Genome Atlas (TCGA). They also used the Keras-vis guided gradient saliency visualization method [19] to identify marker genes for a given cancer. DL has shown considerable potential in using transfer learning with CNNs to predict cancer survival using gene expression data [20]. The authors have proposed a method for rearranging sequence data by transmuting RNA-sequencing samples into gene expression images. These images are then fed into a pre-trained CNN for survival prediction. However, as these methods require molecular functional information for gene rearrangement, it is not possible to use it for other publicly available datasets.

Tumor molecular subtyping, an important step toward personalized treatment, provides crucial biological insights into cancer heterogeneity. The performance of deep cancer subtype classification (DeepCC) [21], an innovative cancer molecular subtype classification framework, has been demonstrated for case studies on colorectal and breast cancer classification. DeepCC is based on functional spectra derived from DL that quantifies the activities of biological pathways. The resistance of DeepCC to missing data has been demonstrated in a simulation based on random subsampling of genes. Furthermore, features learned by DeepCC identify biological properties connected to different molecular subtypes, allowing for more efficient within-subtype and between-subtype patient sample segregation.

[22] have revealed a unique technique for obtaining personalized biomarker scores that explains the importance of each gene in identifying the cancer subtype for each sample by using an attention mechanism. This personalized importance score is then used for a post-classification analysis. PCA and non-negative matrix factorization (NMF) were performed to detect emerging patterns within the biomarker scores. However, as PCA and NMF are independent of the classification task, therefore, there is a need of an end-to-end approach that can be directly employed for real-time applications. PET images containing metabolic information have also been used for the sub-classification of the lung cancer [23]. However, it is difficult to extract enormous PET images of carcinoma patients for deep learning based applications.

### *1.3 Our proposal*

The existing DL classification models inhibit the processing of an entire set of genes and mainly rely on executing feature selection as a prerequisite task before training the downstream classifiers[24-26]. Gene feature subset selection assists in the reduction of redundant features in a dataset that is skewed because of a very high feature-to-sample ratio. However, feature selection has some limitations. First, feature selection approaches are computationally expensive because they use an embedded classifier to evaluate the performance. Furthermore, feature extraction is independent of the classification task; therefore, the downstream classifiers do not have the opportunity to discover hidden information besides the omitted genes and their influence on the respective tumor [27, 28]. Advancements in natural language processing (NLP) offer the opportunity to generate a dynamic representation of features from self-attention based architectures, in particular Transformers [29]. As a result, the global knowledge of a network, attained from a distinctive class, is boosted by the local knowledge that each feature provides. Unlike the traditional gene expression classification techniques, the attention mechanism sequentially selects subsets of input genes and consequently reveals a set of scores that define the significance of each gene for classifying subtypes [30], that is, it emphasizes only the genes that are relevant to a task.

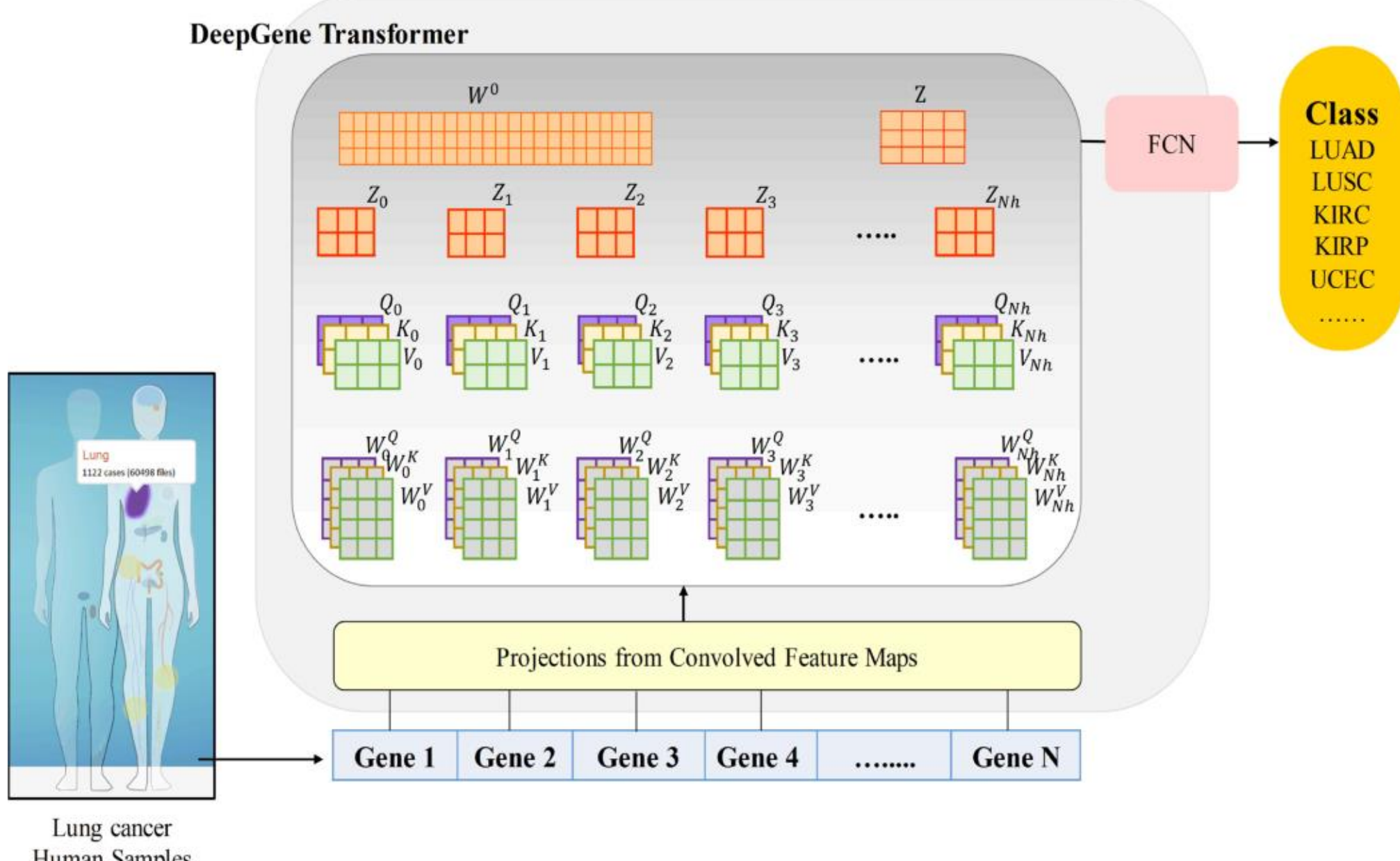

**Figure 1: Graphical representation of the end-to-end deep learning approach for the classification of the cancer subtypes.**

In this study, we proposed an end-to-end approach for lung cancer subtype classification using a gene expression dataset, wherein a multi-head self-attention module permitted the model to jointly learn complex genomic information from thousands of genes from different patient samples shared across multiple cancer subtypes. Head collaboration achieves better generalizability over imbalanced datasets and produces a more desirable performance for both binary and multiclass classification tasks.

The main contributions of this study are as follows:

- DeepGene Transformer architecture is inspired from Transformer encoder architecture that uses the concept of a multi-head self-attention mechanism with one-dimensional (1D) convolution layers as a hybrid architecture to assess high-dimensional gene expression datasets and explore whether the representation learned from the attention mechanism can achieve better performance than the established approaches.

- Our work is among the few studies [31, 32] to suggest that a multi-head self-attention layer with an adequate number of heads can perform 1D convolutions and is less expensive than ordinary 2D convolutional layers.

Inspired by the transformer success in NLP, multiple works tried combining CNN-like architectures with self-attention [33, 34]. To our knowledge, DeepGene Transformer is among the preliminary efforts to develop a comprehensive classifier that does not require gene feature selection as a prerequisite for cancer subtype classification using multi-omics data from different cancer patients.

## I. MATERIAL AND METHODS

### *A. Tumor gene expression dataset*

TCGA [35] is a landmark cancer genomics consortium comprising mRNA expression, imaging, DNA methylation, copy number variations, somatic mutations, and protein expression data. TCGA comprises sequenced and molecularly characterized expression data of over 20,000 primary cancers and matched non-diseased samples from 33 cancer types.

Each step in the genome characterization pipeline generates various data points, such as clinical information and molecular analytic metadata. In this study, we used RNA-sequencing values from lung adenocarcinoma (LUAD) and lung squamous cell carcinoma (LUSC) datasets for classifying lung cancer subtypes. Gene expression values were represented by log2(FPKM + 1), where FPKM is the number of fragments per kilobase per million mapped reads. The number of cancerous and healthy samples for each lung cancer type are outlined in Figure S1 of supplementary file.

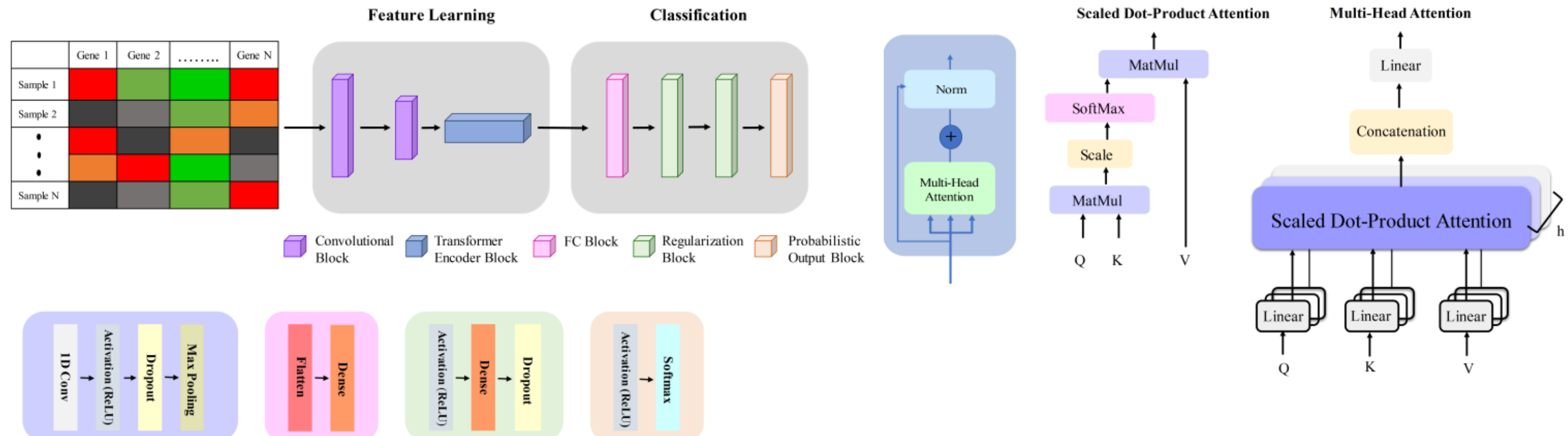

**Figure 2. DeepGene Transformer network: an illustration. (a) (left) Detailed description of DeepGene Transformer architecture. The model includes two parts: feature learning block and a fully connected block for classification. One-dimensional (1D) vectorized gene expression levels from The Cancer Genome Atlas (TCGA) samples were used as the input. This input was then convolved and fed into the multi-head self-attention module for identifying relevant biomarkers. (b) (Right) Scaled dot-product attention to compute the attention function on a set of queries and multi-head attention to concatenate the yielded output values and project them for downstream tasks. The exhibition of the Transformer encoder was inspired by [36].**

## *B. Preprocessing*

Preprocessing is a significant process for appropriately analyzing input data. Preprocessing tasks include the exclusion of categorical rows and columns, exclusion of constant and quasi-constant features from samples, and converting the data into a numeric feature matrix. Here, we eradicated categorical rows (e.g., gene IDs) and categorical columns (e.g., class labels) from the raw input data. We then generated labels from the raw data and stored them in different variables. Unlike previous studies, we did not examine cancer-specific biomarkers to train the model, as we aimed to develop a generalized classifier. After searching for missing values, we fractioned the entire dataset into training, validation, and testing sets.

## *C. Model architecture*

Transformer [36] is a prominent deep learning architecture which was originally proposed for machine translation and now has been widely adopted in various fields such as computer vision, audio processing and multimodal applications. In this line of work, Transformer encoder is used as DeepGene Transformer backbone architecture. The outputs of the encoder are utilized as a representation of the input and used for the downstream task such as classification.

A key challenge for applying Transformer is it inefficiency of processing long sequences mainly due to the computational and memory complexity of the self-attention module. Assuming that the hidden dimension $D_m$ of the model is D, table 1 illustrates the computational complexity and the number of parameters requirement for the Transformer.

| Module | Complexity | # of Parameters |
|---|---|---|
| Self-attention | $O(T^2 \cdot D)$ | $4D^2$ |
| Position wise FFN | $O(T \cdot D^2)$ | $8D^2$ |

**Table 1. Complexity and number of parameters for self-attention and position wise FFN module.**

When the input sequences are long, the sequence length $T$ gradually dominates the complexity of these modules, in which case self-attention becomes the bottleneck of Transformer. Furthermore, the computation of self-attention requires that a $T \times T$ attention distribution matrix is stored, which makes the computation of Transformer infeasible for long-sequence scenarios (e.g., long text documents and pixel-level modeling of high-resolution images) [37].

To deal with this problem, we used two 1D convolutional layers before the multi-head self-attention module in the encoder.

Convolutional layers capture low and high level features, and then feed convolved feature maps to the multi-head self-attention module to jointly attend the information from different representation subspaces.

Different CNN models have been proposed for cancer subtype classification. Generally, a model accepts a 2D matrix, such as an image as an input; 2D kernels are used to extract local features in the input space. However, in a gene expression dataset, each patient sample is oriented in a 1D array; therefore, traditional CNN may not be a viable approach for tumor classification. Alternatively, an improvised version of 2D CNN known as 1D CNN has been recently proposed. 1D CNNs are advantageous and have instantly achieved state-of-the-art performance in numerous applications owing to the following reasons:

- In a 2D CNN, an image with N × N dimensions convolves with K × K kernel ~$O(N^2K^2)$, whereas in a 1D convolution with similar dimensions, N and K, this is ~O(NK). This indicates that under similar conditions, a 1D CNN is significantly less computationally expensive than a 2D CNN [38].

- The majority of 1D CNN applications use one or two hidden layer network configurations with <10,000 parameters, whereas all 2D CNN architectures use deeper architectures with more than 1 M (usually above 10 M) parameters. As a result, 1D CNNs are easier to train and are ideal for real-time applications.

- Shallower models are frequently preferable for basic applications where the number of samples is limited relative to the number of features, such as in cancer subtype classification where a compact architecture of 1D CNN is used. Such shallow networks avoid overfitting and require fewer training resources.

Considering such characteristics of 1D CNNs, we attempted to develop a novel and generalizable DL framework that leverages a 1D CNN and multi-head self-attention mechanism for cancer subtype classification without performing feature selection as a prerequisite task. Figures 1 and 2 show the proposed DL architecture for tumor gene expression in further detail. A graphical illustration of the proposed end-to-end learning system is shown in Figure 1, and the layer-specific details are depicted in Figure 2. The first layer is the input layer, which takes the gene expression data as a vector. The second layer is a convolutional block consisting of a 1D convolutional layer, a nonlinear activation function (ReLU), and an intermediate max-pooling layer. The max-pooling layer had a pool size of one and stride size of one with zero padding. A 1D convolutional layer in the convolutional block consists of a set of kernels that locomotes across input vectors and detect certain characteristics in any position on the input feature map. Considering the convolutional operation in the 1D convolution layer, the computation can be written as:

$$X_j^{(l)} = f\left(\sum_{i=1}^{M} X_i^{(l-1)} * W_j^{(l)} + b_j^{(l)}\right) \quad (1)$$

where j indicates the size of the kernels, and M refers to the channel numbers of the input $X_i^{(l-1)}$. $W_j^{(l)}$ is the j-th convolutional kernel corresponding to the M-th channel, and $b_j^{(l)}$ is the kernel bias corresponding to the j-th kernel. The activation function is denoted as *f*, and * is the element-wise multiplication.

A 1D convolutional layer, a drop-out layer, and a max-pooling layer constituted the second convolutional block. The drop-out layer was introduced to avoid overfitting. The max-pooling layer selected the significant features in the default window and sped up the process by down-sampling the number of parameters. In this block, the pooling layer had a pool size of 10 and a stride of one with no paddings.

In the multi-head self-attention module, the feature maps from the second convolutional blocks were segregated into a fixed number (h, number of heads) of segments, and then self-attention was applied to the corresponding segment, resulting in a h context vector for each gene. The final context vector was obtained by concatenating all h context vectors. The attention was calculated from three key components: Q (query), K (key), and V (value). When the Q, K, and V are all obtained from an identical input sequence X, it is called self-attention. The Q and transpose of K were passed to a scalar dot product block to calculate the attention weights on the values. The obtained weights were then passed through a softmax activation function for dimension normalization, and weighed in combination with respective values to achieve the final attention. These weights were then concatenated and projected once more, yielding the final values as shown in Figure 2.

Because each multilayer perceptron flattens the output space, this generates a vector containing the attention scores of specific genes, which is extremely accountable for a particular cancer subtype. Next, this vector was infiltrated through a fully connected module and regularization modules, and a softmax layer for determining the probability distribution over the classes. Algorithm 1 depicts the pseudocode for computing the feature importance through the attention mechanism for the classification of the lung cancer subtypes

### *D. The multi-head self-attention layer*

Let X ∈ $R^{T\times D_{in}}$ be an input matrix consisting of T tokens in $D_{in}$ dimensions of the raw feature space. In NLP, each token represents a word in a sentence; the same formalism can be applied to sequences of T discrete objects such as pixels [36]. In the case of the 1D convolutional layer, the input was a vector of gene expression values, where each token corresponded to the patient sample expression values for a particular gene.

A self-attention layer maps any query token $t \in [T]$ from $D_{in}$ to $D_{out}$ dimensions, as follows:

$$Self-Attention(\boldsymbol{X})_{t,:} := Soft\max(\boldsymbol{A}_{t,:})X\boldsymbol{W}_{val} \quad (2)$$

$$A := \boldsymbol{X}\boldsymbol{W}_{qry}\boldsymbol{W}_{key}^{T}\boldsymbol{X}^{T} \quad (3)$$

as attention scores, and softmax output as attention probabilities. The layer was parameterized by a query matrix $\boldsymbol{W}_{qry} \in R^{D_{in}\times D_k}$, a key matrix $\boldsymbol{W}_{key} \in R^{D_{in}\times D_k}$ and a value matrix $\boldsymbol{W}_{val} \in R^{D_{in}\times D_{out}}$.

To keep equation (2) consistent for the 1D layer, we used the notation of a slice tensor by utilizing a 2D indexed vector: if r = (i, j), we represented it as $X_{r,:}$ and $A_{r,:}$ changed to mean $X_{i,j:}$ and $A_{i,j,:,:,}$ respectively. In lieu of this notation, the multi-head self-attention output at gene element Q can be stated as:

$$Self-Attention(\boldsymbol{X})_{Q,:} := \sum_{\boldsymbol{K}} Soft\max(\boldsymbol{A}_{Q,:})_{\boldsymbol{K}} \boldsymbol{X}_{\boldsymbol{K},;}\boldsymbol{W}_{val} \quad (4)$$

Self-attention is equivariant to reordering, which means that it produces identical results regardless of how the T input tokens, matrix, or vector values are shuffled. This is challenging in situations where the order of events is expected to be important. Before applying self-attention, a positional encoding for each token in the sequence (or pixel in an image) is learned and added to the representation of the token itself. However, for a vectorized input from a gene expression dataset, positional encoding is not required.

As the multi-head attention module transmutes Q, K, and V into the h subspace with distinct learnable linear projections, it is advantageous to replicate the self-attention mechanism into several heads rather than conduct a single attention function with $D_{model}$ -dimensional keys, values, and queries:

$$Q_h, K_{h,} V_h = Q\boldsymbol{W}_h^Q, K\boldsymbol{W}_h^K, V\boldsymbol{W}_h^V \quad (5)$$

where $\{\boldsymbol{Q}_h, \boldsymbol{K}_h, \boldsymbol{V}_h\}$ are the query, key, and value representations of the h-th head, respectively.

The reason behind capturing diverse context features with multiple individual self-attention functions is that each head focuses on distinct parts of a vectorized input by using the query, key, and value representations, and all relevant information is jointly attended to. The outputs of the $N_h$ heads of the output dimension $D_h$ are concatenated and projected to dimension $D_{out}$ in multi-head self-attention (MHSA) as follows:

$$MHSA(X) = concat_{h\in[N_h]}[head_1,....,head_h]\boldsymbol{W}_{out} + \boldsymbol{b}_{out} \quad (6)$$

Where,

$$head_i = Self-Attention(Q\boldsymbol{W}_{h_i}^Q, K\boldsymbol{W}_{h_i}^K, V\boldsymbol{W}_{h_i}^V)$$

**Remark 1**: MHSA as a convolutional layer
**Theorem 1**: In a model, a multi-head self-attention layer with $N_h$ heads of dimensions $D_h$ and output dimensions $D_{out}$ acts as a convolutional layer with kernel size $\sqrt{N_h}$ x $\sqrt{N_h}$ and min ($D_h$, $D_{out}$) output channels. Authors in [31] state that multi-head self-attention with $N_h$ heads can simulate a 1D convolutional layer with a kernel of size $K = N_h$. To prove this, we implemented the same formalism in the proposed DeepGene Transformer architecture.

This theorem could be proved by effectively choosing the hyperparameters of the MHSA layer, so that it may act as a convolutional layer.

- Padding: A convolutional layer reduces the image dimension by K-1 pixels, whereas, a MHSA layer employs the "SAME" padding mode, which equally fills four edges of the input of each layer with zero to maintain the same size of the inputs and outputs. The output from the MHSA and convolutional layers would be identical if the input image was padded with K/2 zeros. In the case of 1D convolutions, padding was not required.

- Stride: The stride was set to one to make the MHSA layer act as a convolutional layer. Theorem 1 was well defined in this study for stride as one in the 1D convolutional layer.

**Remark 2**: DeepTRIAGE [22] adopted the concept of self-attention, in which attention to a feature is impacted solely by that feature. In the proposed DeepGene Transformer framework, we modified the self-attention concept in such a way that knowledge was explored by multiple heads, instead of one-by-one. The learned weights for query (Q), key (K), and value (V) were l by $D_K$, l by $D_K$, and l by $D_V$, respectively, with $D_K$ and $D_V$ set to 64 by default.

## II. Results

The proposed architecture was implemented using the "Scikit-learn" and Keras DL platform with a TensorFlow backend in Python. GPU-specific implementation was used. All the tests were run on a PC with an Intel Xeon i7-8700K processor (3.70 GHz), 128 GB of RAM, and an Nvidia GTX 1080 Ti. During training, we experimented with several configurations for common hyperparameters (learning rate, network depth, and drop-out rates) as well as hyperparameters specific to our proposed model (number of heads, attention key, and value dimensions). Table 2 summarizes the comprehensive description of the hyperparameters specific to our model. We did not verify whether reducing the attention key size $D_K$ had an impact on the model quality. Per our observations, the highest training and validation model accuracy was achieved through an epoch size of 95 with different dimensions of the kernel filters, depending on the number of heads for the self-attention layer; however, if the validation accuracy did not improve for five consecutive epochs, the training was stopped early.

Results based on both typical and model-specific hyperparameters produced through random search and 3-fold cross-validation were implemented using a comparative analysis with macro-averaged precision, recall, and F1 score. When a dataset is imbalanced, accuracy provides an overoptimistic approximation of the majority class [39]. Matthew's correlation coefficient (MCC) scores were also reported, which lay in the range of [-1, +1], where MCC = 1 is indicative of perfect classification, MCC = -1 of complete misclassification, and MCC = 0 of random coin tossing prediction.

Five different trials were run while training the model, and the mean accuracy of all assessments was reported for various training and validation data partitioning procedures. To assess the effectiveness of the proposed architecture, we used multiple data splitting strategies for binary (LUAD and LUSC) and multiclass (normal, LUSC, and LUAD) classification. For the binary classification problem, the entire dataset was divided into three parts: 70% for training, 20% for validation, and 10% for testing. However, three different data-partitioning schemes were adopted for multiclass classification problems. The first method divided the data into three parts: 60% for training, 20% for validation, and 20% for testing; the second divided the data as: 70% for training, 20% for validation, and 10% for testing; and the third divided the data as: 80% for training, 10% for validation, and 10% for testing.

The main objective of this study was to propose an end-to-end approach that works reasonably well for both binary and multiclass datasets in comparison with traditional ML algorithms. Typically, gene expression data contain hundreds to thousands of high-dimensional imbalanced patient samples, which reduces the efficiency of data mining approaches because of the problem of overfitting. To deal with the curse of dimensionality, ML models usually employ gene selection/extraction methods to extract disease-mediated genes from redundant data. PCA, recursive feature elimination (RFE), and mutual information methods were analyzed in combination with LR, RF, and SVM, which are the traditional ML models (Table 2).

For each methodology examined here, the mean performance rates of 15 validation subgroups from a 5-repeated 3-fold cross-validation procedure are presented in Table 2. To assess the importance of the proposed DeepGene Transformer framework, we modified the number of heads ($N_h$), as well as the attention key ($D_K$) and value ($D_V$) dimensions, keeping the model dimensions ($d_{model}$) and drop-out ($P_{drop}$) rate constant. The results for the proposed DL model with various configurations are shown in the first two rows of the table, whereas, the performance rates for the traditional ML techniques are shown in the last nine rows. The dashed-dotted lines segregate the ML approaches according to the feature selection strategies employed by the model (PCA, RFE, and MI) to limit the number of irrelevant/correlated features. This comparative analysis was only performed for binary classification.

From the values presented in Table 2, the proposed CNN architecture outperformed all traditional ML approaches with a testing accuracy of 98% with a 4-head and 99% with an 8-head configuration. In particular, the best values for all evaluation metrics were obtained using the proposed CNN with the 8-head configuration. The combination of the PCA feature extraction method and the SVM classifier produced the best results (0.93 area under the curve [AUC]) among the ML techniques studied, followed by mutual information with SVM (0.92 AUC).

Table 3 shows the mean precision, recall, and F1 score for the three alternative data splitting algorithms used for multiclass classification training, validation, and testing with varied configurations. The proposed DeepGene Transformer framework achieved a higher testing accuracy in 80% training and 10% of both validation and testing strategies. The more training data an architecture has, the better the performance is achieved from the proposed framework. A box plot of the distribution of the 5-repeated 3-fold cross-validation (CV) and MCC values obtained by the proposed DL architecture with various configurations, as well as the ML techniques is shown in Figure 3. The same color shades correspond to similar DL and ML approaches with different configurations. The proposed DeepGene Transformer architecture with 4- and 8-head configurations has a 10% higher class-specific MCC score than the conventional ML techniques that use well-established feature selection methods for the selection of significant genes.

The improvement of validation accuracy after the training phase is another important process for evaluating the performance of any DL model. Growth in the validation accuracy demonstrates that the learning process is improving. Figure 4 shows the progress of validation accuracy and loss during the training process for the binary classification of LUAD and LUSC with 4 head configurations. The accuracy curves generated by the proposed DeepGene Transformer framework in Figures 4 (a) shows that the accuracy was stable across the CV folds, implying that the predictions of the proposed DeepGene Transformer framework were far superior to those by random guessing. Figures 4 (b) depicts the progression of the validation accuracy through the training process for multiclass classification between the normal, LUSC, and LUAD patient samples, with 4 head configurations. The progress of accuracy and cross entropic loss during the training and validation phase for binary and multiclass classifications with 8-head configurations are shown in Fig S2 (a & b), provided as Supplementary File.

A related study [40] has demonstrated a similar contribution of the same dataset using a microarray-gene set reduction algorithm. Previous studies have employed extra integrative correlation filtering as a feature selection method to eliminate heterogeneity. Four different evaluation metrics were used in this study; however, we only used the area under the precision-recall curve (AUPR) and generalized Brier score (GBS) to prove the effectiveness of DeepGene Transformer for tumor classification. GBS is the mean squared difference between the predicted probabilities and observed outcomes and is used for the verification of the probability forecast [41]. AUPR is an estimate of averaged precision across all recall values and is observed as a measure of model uncertainty for imbalanced or skewed datasets [42]. As AUPR gets closer to 1, the classifier performs better. In contrast, lower GBS values are more desirable. Table 4 clearly illustrates that the proposed architecture outperformed architectures in related studies in terms of evaluation metrics.

## III. Discussion

The multi-head attention mechanism has become a key component of DL networks, contributing to achieving superior results in machine translation and other NLP tasks [43]. Beyond performance and computational cost, RNNs have been used in conjunction with attention mechanisms; attention mechanisms have also gained traction in computer vision and bioinformatics. Motivated by these studies, we developed a MHSA mechanism for the 1D CNN model by encouraging individual attention heads to extract distinct information for the prediction of lung cancer subtypes.

Transcriptomic subtypes of cancer are usually classified by selecting the expression levels of a subset of genes associated with the progression and differentiation of tumor cells [44, 45]. However, it is unclear whether the dysregulation of the selected subset of genes affects the exact morphology of the tumor tissues. To address this problem, we proposed an end-to-end approach that utilizes all the expression values for classification.

By developing DeepGene Transformer, we have extended the MHSA mechanism to some high-dimensional imbalanced gene expression dataset, where there are thousands of gene features for a few hundred patient samples. In a typical attention layer, once the model has been trained, we use softmax to activate the weights of the attention layer. With only a few convolutional blocks without attention mechanism, in the DeepGene Transformer architecture, the model collapsed while training with a high-dimensional gene expression dataset. In the proposed DeepGene Transformer architecture, the self-attention acts as a convolution layer under sufficient conditions that sparsifies the input space, emphasizing only the relevant genes and filtering the remainder. Multi-head aggregates the information from different attention heads to generate a more biologically meaningful feature representation for the downstream classification.

## IV. CONCLUSION

In 2018, lung carcinoma had the highest incidence (2.1 million new cases) and mortality (1.8 million deaths) worldwide. Lung carcinoma is a heterogeneous disorder with several distinct subtypes. Since 2003, more than 20 lung cancer risk prediction models have been published, and all of them use feature selection as a prerequisite for subtype classification. However, some models are application-specific, that is, they only work well for a few classes and balanced datasets. Our framework, DeepGene Transformer is an end-to-end approach that prioritizes features during testing and outperforms the existing state-of-the-art methods in the case of both binary and multiclass problems. Based on the experimental results, we concluded that fusion of the MHSA mechanism with 1D CNN supported high-dimensional microarray datasets without any computational complexity. The DeepGene Transformer approach is a preliminary attempt to investigate how attention mechanisms can be used to predict lung cancer subtypes. Our findings indicate that employing the attention mechanism can help researchers better understand the relationship between patient samples and gene expression data. We also compared the results of DeepGene Transformer with those of traditional ML methods. Based on various evaluation metrics, we demonstrated that the proposed framework outperformed the current state-of-the-art frameworks. This approach can be expanded in the future to define subtypes of other cancer types, allowing for comprehensive cancer prognostic research. Moreover, it can be combined with histopathological parameters to reveal “omics” features associated with histopathology and identify morphological variations associated with gene expression.